\documentclass[onecolumn,12pt]{IEEEtran}
\usepackage{graphicx}
\usepackage{amsmath}
\usepackage{amssymb}
\usepackage{cite}
\usepackage{color}
\usepackage{xcolor,colortbl}
\usepackage{atbegshi}
\AtBeginDocument{\AtBeginShipoutNext{\AtBeginShipoutDiscard}}

\begin{document}
\title{Stable Self-Assembled Atomic-Switch Networks for Neuromorphic Applications}
\author{Saurabh K. Bose, Joshua B. Mallinson, Rodrigo M. Gazoni, and Simon A. Brown}
\thanks{The authors are with The MacDiarmid Institute for Advanced Materials and
Nanotechnology, Department of Physics and Astronomy, University of
Canterbury, Private Bag 4800, Christchurch 8140, New Zealand
(email:simon.brown@canterbury.ac.nz)
\\The authors gratefully acknowledge financial support from the
Marsden Fund, New Zealand, and the MacDiarmid Institute for
Advanced Materials and Nanotechnology. 
\\ This is a post-peer-review, pre-copyedit version of an article published in IEEE Trans. Elect. Dev. 2017. The final authenticated version is available online at: http://dx.doi.org/10.1109/TED.2017.2766063”.}
\maketitle

\begin{abstract}
Nature inspired neuromorphic architectures are being explored as
an alternative to imminent limitations of conventional
complementary metal-oxide semiconductor (CMOS) architectures.
Utilization of such architectures for practical applications like
advanced pattern recognition tasks will require synaptic
connections that are both reconfigurable and stable. Here, we
report realization of stable atomic-switch networks (ASN), with
inherent complex connectivity, self-assembled from percolating
metal nanoparticles (NPs). The device conductance reflects the
configuration of synapses which can be modulated via voltage
stimulus. By controlling Relative Humidity (RH) and oxygen
partial-pressure during NP deposition we obtain stochastic
conductance switching that is stable over several months. Detailed
characterization reveals signatures of electric-field induced
atomic-wire formation within the tunnel-gaps of the
\textit{oxidized} percolating network. Finally we show that the
synaptic structure can be reconfigured by stimulating at different
repetition rates, which can be utilized as short-term to long-term
memory conversion. This demonstration of stable stochastic
switching in ASNs provides a promising route to hardware
implementation of biological neuronal models and, as an example, we highlight possible applications in Reservoir Computing (RC).

\end{abstract}


\section{Introduction}
\IEEEPARstart{T}{he} astounding success of the von Neumann
architecture for computers\cite{Goldstine1946}, as encapsulated in
Moore's Law, is now meeting with fundamental limitations (physical
transistor dimensions are approaching classical limits) and
practical limitations (the exponential increase in research and
development costs for every new process
line)\cite{Taur1997,Frank2002}. Natural information processing
systems, like the biological brain, on the other hand, can perform
highly complex computational tasks like  navigation, recognition
and decision-making with remarkable ease and with very low energy
consumption\cite{Kandel2000}. This natural computation, processing
the useful data (patterns) from a multitude of sensory
information, is immediate and cannot be matched by even the
most-advanced supercomputers\cite{Roska2005,Wong2012}. Nature
inspired
architectures\cite{Tuma2016,Bose2015,Service2014,Avizienis2012,
Stieg2012, ohno2011, Suri2013} are therefore currently being
pursued as a disruptive alternative to the von Neumann
architecture. A recent review on Neuromorphic
architecture\cite{Mead1990a} and implementations can be found in
Ref. \cite{Nawrocki2016}.

\begin{figure}[!t]
\centering
\includegraphics[width=10 cm ]{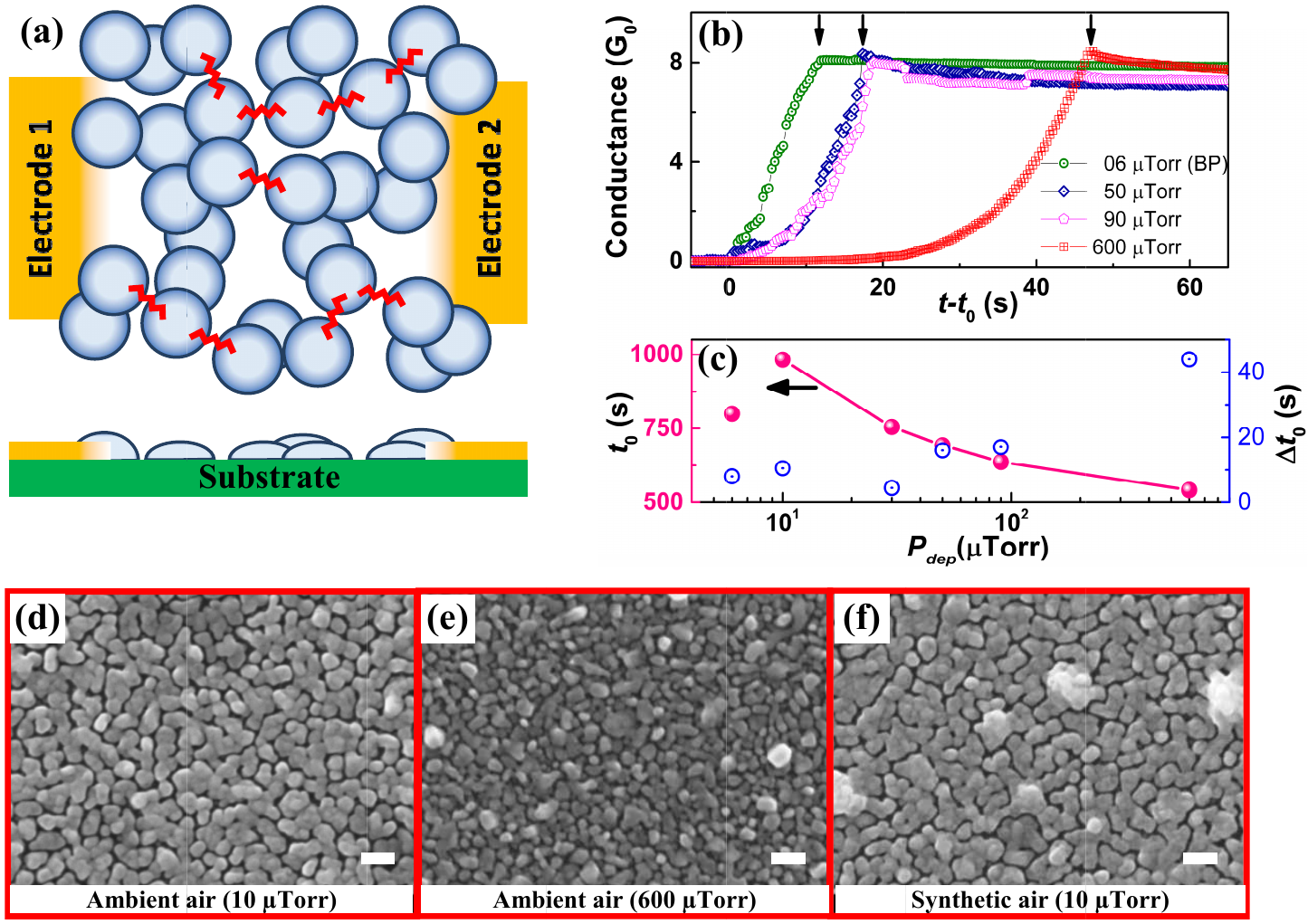}
\caption{(a) Schematic depiction (top and side view) of the
nanoparticle network  between electrical contacts of our two-electrode devices showing
tunnelling gaps in the Sn NP network. The outer shaded region on
the NPs depicts the thin oxide layer. (b) Variation of the onset
of conductance with increase in partial pressure $P_{dep}$ (over 2
orders of magnitude) during NP deposition. (c) This pressure
variation results in shorter conduction onset times $t_0$ (left
scale) with longer onset width $\triangle t_0$ (right scale). (d)
Scanning electron micrograph of samples prepared at lowest ambient
air pressure ($P_{dep}\sim$10 $\mu$Torr) shows more coalescence
and larger grain-size in comparison to the highest pressure (600
$\mu$Torr) prepared samples shown in (e). (f) Use of dry synthetic
air ($P_{dep}\sim$10 $\mu$Torr) produces a different
microstructure with reduced sample stability as compared with
ambient air. The white scale bars are 100 nm.}

\label{bose1}

\end{figure}

The alternative brain-inspired hardware approach must address
three key issues simultaneously: mimic the complex biological
network of neurons, replicate synaptic structures and allow
implementation of standard computational
algorithms\cite{Sterratt2011}. Achieving all of these goals is obviously enormously
challenging and will require long-term research. Nevertheless significant progress
has been made towards solving several different problems, using a variety of architectures.
Proposals for non-CMOS approaches include those based on networks of memristors\cite{Indiveri2013,Querlioz2013a,Adam2017},
atomic switches\cite{Avizienis2012,Stieg2012} and Metal Oxide Resistive
Random Access Memory (RRAM)\cite{Yu2011,Park2016}. There have been interesting demonstrations
of memristor-based neural networks\cite{Thomas2013}, associative
memory\cite{Pershin2010}, synaptic emulators\cite{Wang2017},
conditional programming\cite{Borghetti2009}, reconfigurable
logic\cite{Xia2009}, solving mazes\cite{Pershin2011}, pattern
recognition\cite{Snider2007, Alibart2013,Chu2015,Sheridan2014},
and reservoir computing
(RC)\cite{Konkoli2014,Kulkarni2012a,Sillin2013}. Even in the most heavily explored architectures
(regular cross-bar arrays of memristors)~\cite{Burr2017} there remain unsolved challenges in regard
to realisation of the required properties of both individual switching elements and networks of these elements.

RC is simpler to implement than many other unconventional computation schemes since synapses do not need to be addressed individually and can  be seen as an important step towards achieving other types brain-like computation. In RC a `reservoir' comprising a complex network of switching elements allows the transformation of input signals into a higher dimensional space.\cite{Konkoli2014,Kulkarni2012a,Sillin2013}  Training of a single `output layer' then allows implementation of various time series prediction, pattern recognition and classification tasks.~\cite{Fernando2003,Choi2015,Demis2016} Randomly assembled atomic switch networks (ASNs) based on sulphidised Ag nanowires~\cite{Avizienis2012,Stieg2012} and percolating films of nanoparticles~\cite{Sattar2013, Fostner2015} are immediately ammenable to RC. ASNs are also an appealing alternative to regular arrays of devices because they allow realisation of complexity  similar to that of the brain and fabrication via self-assembly immediately circumvent the limitations of lithographic processing. Ag-based ASNs have recently been used to demonstrate a form of RC in which the non-linear properties of the reservoir allows generation of target waveforms, and a clear roadmap towards further implementation has been mapped out\cite{Demis2016}.

Systems of inorganic synapses are however in the early stages of
development with improvements required in production methods, reliability and actual functionality\cite{Burr2017}. While robust switching over 10,000 cycles has been reported for Ag-AgS nanowire systems\cite{Avizienis2012} a different, and a particularly important, issue for any real-world applications is long-term device stability, which is the main focus of the present work. Such stability has not been reported previously in either Ag-AgS nanowire \cite{Avizienis2012,Stieg2012,Sillin2013,Stieg2014,Demis2016} or percolating ASNs\cite{Sattar2013}.

Here we report a straightforward fabrication procedure for
realization of randomly connected ASNs within a percolating network of metal
nanoparticles (NPs). We show that deliberate introduction of
oxygen and moisture during NP deposition leads to long term device
stability. Despite the presence of oxides, the switching
mechanisms associated with increases(decreases) in device
conductance $G_{\uparrow}$($G_{\downarrow}$) are shown to be
formation (destruction) of atomic scale wires in tunnel gaps in
the network. The two-terminal device conductance (\emph{G})
quantifies the input-output electrode
connectivity\cite{Fostner2015} and reflects the \emph{synaptic
configuration} of the network which can be reconfigured by voltage
stimulation. Finally, we discuss the observed synaptic
stochasticity and why it is useful for implementation of hardware
analogues of the biological brain\cite{Rolls2010}.

\section{Experimental}

The nano-cluster deposition system used in this study is based on magnetron sputtering to generate a vapour of the metal of interest and gas aggregation to condense the vapour into particles, and has been described in detail in previous publications\cite{reichel2006a,Dunbar2006,Ayesh2007}. The deposition scheme provides a narrow cluster size distribution\cite{reichel2006a} and allows precise control over the NP surface coverage near the percolation threshold\cite{stauffer}, so that the system is poised near criticality\cite{Chialvo2010}. Sn NPs with mean diameter $\sim$ 8.5 nm are deposited between 50 nm thick Au/NiCr electrodes on Si$_3$N$_4$ substrates, with active area of 100 $\mu$m $\times$ 300 $\mu$m. The two-contact devices allow for a demonstration of network stability and associated dynamics, but samples with multiple contacts will be required for demonstration of RC.

The Sn NPs are deposited at room temperature which
means that ordinarily the surface atoms have sufficient mobility
to allow coalescence\cite{Yu1991}. For samples poised near the
percolation threshold, the coalescence can lead to the loss of
conducting pathways through the film, because neighboring
particles that are initially joined by a fragile connection are
pulled apart as they coalesce with other neighbor NPs, thus
contributing to the short life-span ($\sim$ hours) of previous
devices\cite{Sattar2013}. In the present work, the coalescence of
the Sn NPs is controlled by partial oxidation (during NP
deposition) via a controlled leak of air with a needle-valve. As
will be shown, the controlled oxide formation leads to reduced
coalescence and enhanced device stability. We emphasize that by
`stability' we do not mean that the device has a fixed
conductance, but that the device is in a state in which it
continues to exhibit multiple switching events in response to
voltage stimuli.

\begin{figure}[!t]
\centering
\includegraphics[width=10 cm ]{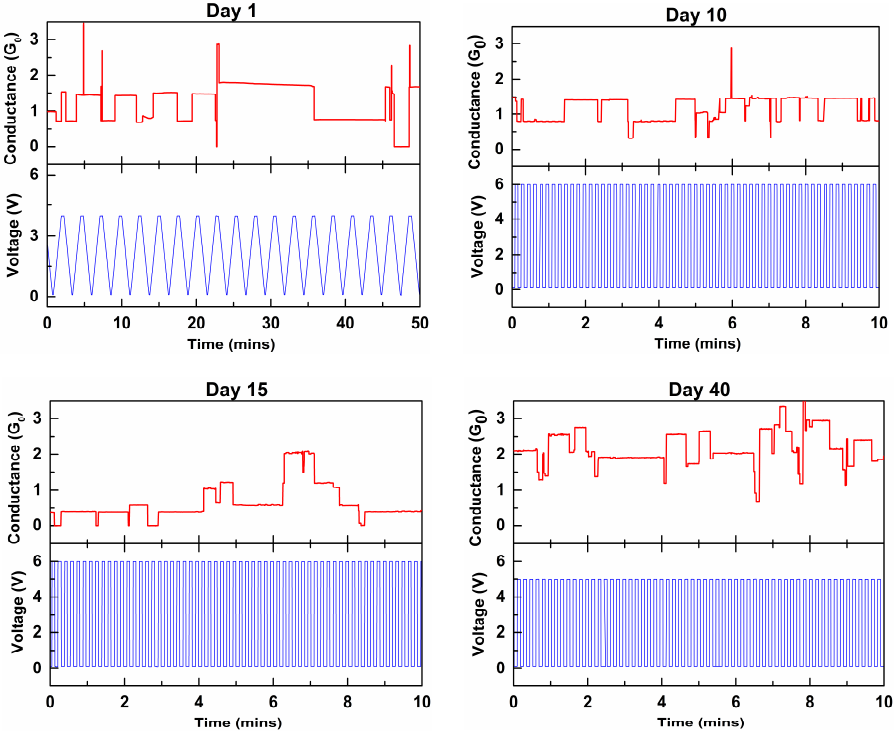}
\caption {Stochastic and stable switching behaviour for sample
prepared with $P_{dep} =$ 10 $\mu$Torr  and high humidity (RH
$\sim$ 80\%) ambient air conditions, showing multi-level
conductance switching, induced by application of both triangular
and pulsed voltage stimulus. On the 1$^{st}$ day, immediately after sample fabrication, we use voltage-sweeps in order to check the voltage threshold for switching, as discussed in section \ref{Dynamics}. On subsequent days we utilize controlled voltage pulses. As we tested the samples in a variety of ways at different times over a period of several months, the applied voltage in panel (d) is slightly different to that in the other panels. The sample exhibits qualitatively similar switching for several months of device operation.}

\label{bose2}

\end{figure}

\section{Sample Fabrication}

\subsection{Self-assembly of ASNs}

We have self-assembled interconnected and active network of
atomic-switches as depicted in the schematic shown in
Fig.~\ref{bose1}(a). Deposition of the Sn NPs at Base Pressure
(BP, $\sim 6 \mu$Torr) led to initial observation of a non-zero
conductance (time $\emph{t}_0$) at around 800s with a sharp onset
behaviour ($\triangle t_0$) $\sim$ 10s, as shown in Fig. 1(b).
Here, we define the width of the onset $\triangle$t$_0$ as the
time after \emph{t}$_0$ required to reach a conductance of 6$G_0$
($G_0 = 2e^2/h $ is the quantum of conductance\cite{VanWees1988}).
The cluster deposition is stopped [arrows in Fig.~\ref{bose1}(b)] when $G\sim$8$G_0$, which represents a nanoparticle surface coverage slightly greater than the percolation threshold, and has been found experimentally to yield optimal
switching behaviour when $P_{dep}$ is in the range $\sim 10-50 \mu$Torr, as described below .
As can be seen in Fig. \ref{bose1}(c) the increase in deposition
pressure ($P_{dep}$) leads to a monotonic \textit{decrease} in
\emph{t}$_0$ coupled with an \textit{increase} in
$\triangle$\emph{t}$_0$. The smooth conduction onset and longer
$\triangle$\emph{t}$_0$ is in contrast to the step-wise conduction
onset of samples deposited at BP (minimum oxidation) in
Ref.\cite{Sattar2013}; those samples were dominated by few
quantized conduction pathways and therefore lacked the large-scale
distributed synaptic network essential for neuromorphic
applications. The post-deposition scanning electron micrographs
(SEM) shown in Fig.~\ref{bose1}(d-e) reveal markedly smaller NP
sizes for devices prepared at higher $P_{dep}$. This can be
understood in the framework of diffusion of the tin NPs on the
substrate being inhibited by the formation of tin-oxide-shell,
leading to reduced coalescence at higher $P_{dep}$. Such formation
of oxide-shell is known to inhibit grain-rotation-induced grain
coalescence (GRIGC)~\cite{MOldovan2002} thus favoring smaller
grain-sizes. This reduced coalescence means that the percolation
threshold is reached more quickly (smaller \emph{t}$_0$) and the
conductance then increases more slowly (larger $\triangle t_0$).
After the deposition is stopped [arrows in Fig.~\ref{bose1}(b)]
the slow conductance change is primarily due a small amount of
further coalescence of the NPs, decreasing $G$.

\begin{figure}[!t]
\centering
\includegraphics[width=10 cm ]{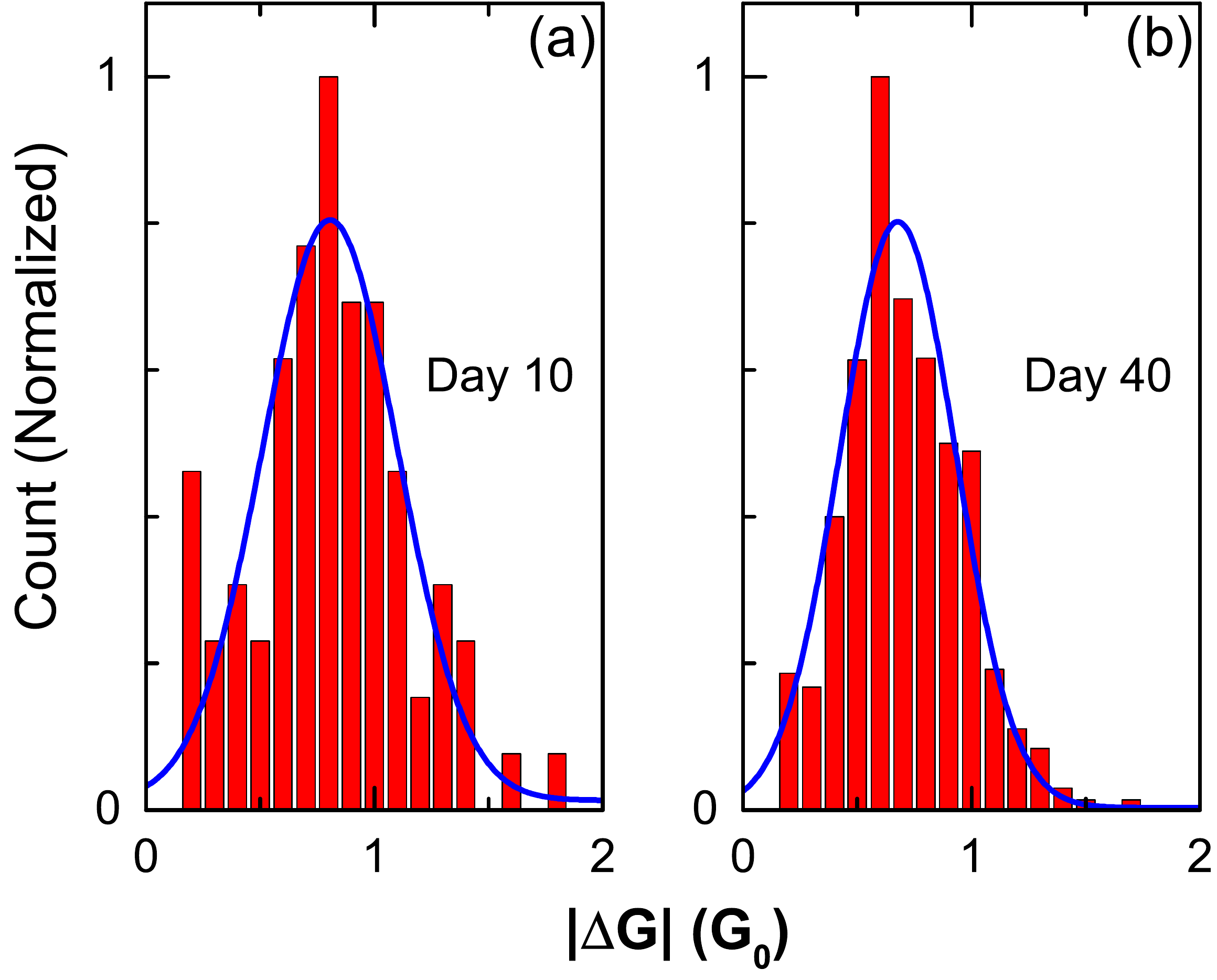}
\caption {The normalized distribution of the change in conductance for the switching events ($\lvert\Delta G\rvert$)
on day 10 and day 40, for long periods of time, shows the stability of the switching dynamics. The solid lines are Gaussian fits to the data. }
\label{bose3}

\end{figure}

\subsection{Optimization of Pressure and Humidity}
The samples fabricated with \emph{high} Relative Humidity (RH
$\sim$ 80\%) ambient air have been stimulated with voltage sweeps
and square voltage pulses over several months. Data for a typical
sample are shown in Fig.~\ref{bose2}. The four panels show
representative snapshots of the switching events on the 1$^{st}$,
10$^{th}$, 15$^{th}$ and 40$^{th}$ day. As described below, the detailed switching behaviour of the network is a complex function of applied voltage, pulse widths ($\tau_p$) and history of the inputs, but qualitatively similar conductance switching is observed for long periods of time. Application of voltage
sweep or pulses induces Electric Field Induced Evaporation (EFIE)
and Electric Field Induced Surface Diffusion (EFISD) of the
surface atoms\cite{Olsen2012}, resulting in atomic-wire formation
in tunnel gaps in the percolating network (resulting in
$G_{\uparrow}$). The electronic flow causes electromigration
induced opening of the previously connected atomic-wires
\cite{Xiang2009} resulting in $G_{\downarrow}$. The conductance
thus switches between multiple conductance states with $G \sim$
1--3$G_0$.  $G$ $\rightarrow$ 0 at multiple points, but the
electric field induces reconnections and results in the network
configuration returning to the connected regime of non-zero
conductance. Samples fabricated in high humidity conditions (RH
$\sim$ 80\%) with ambient air have been tested in this way for
periods of several months without the sample becoming permanently
open circuit. The key point is that the samples prepared with high
RH oxidation exhibit stable switching behaviour without
significant performance degradation. For example, Fig.~\ref{bose3}
shows the distribution of the change in conductance for switching events ($\lvert\Delta G\rvert$)
on day 10 and 40 (see Fig.~\ref{bose2}). Both the mean and variance of the distributions are very similar, indicating that the average switching behavior is the same. Obviously it is an onerous task to test such samples for much longer periods and further testing is required to determine the ultimate lifetime of the samples.Both the mean and variance of the distributions are very similar.
Obviously it is an onerous task to test such samples for much longer periods and further testing is
required to determine the ultimate lifetime of the samples. The same kind of stable switching behaviour is observed for voltages upto ($\geq$ 7-8V) but application of very high voltages ($\geq$ 10V) causes irreversible breakdown in the devices.

\begin{figure}[!t]
\centering
\includegraphics[width=10 cm ]{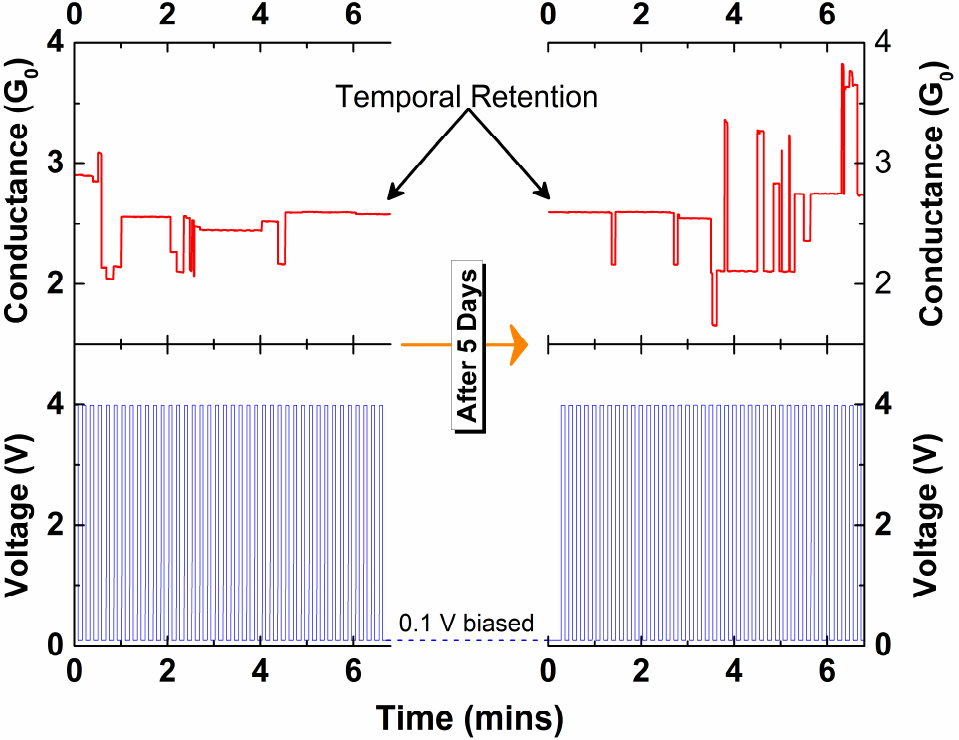}
\caption {Temporal retention of the network conductance tested over more than one week.
Left: the device shows switching in response to 4V pulses before a constant `read' voltage (0.1V) was applied for 5 days.
Right: the conductance showed no measurable change and
switching re-commenced on application of 4V pulses again. }

\label{bose4}

\end{figure}

The oxidation of pure Sn into tin-oxides [SnO, SnO$_2$] is well
known to be accelerated under humid conditions.\cite{Cho2005} Studies of oxidation of Sn in conditions similar to the present ones show that only partial surface oxides are formed.\cite{Sutter2014,Lee1998} Ex-situ analysis of the present oxide-structure is obviously not feasible and so instead we have investigated in-situ the device stability over several weeks. Fig.~\ref{bose4} shows that the device conductance did not change measurably when the device was left for 5 days with 0.1V applied. The conductance switching is resumed when voltage pulses were applied again, which clearly indicates that the oxide structure did not evolve significantly in this period.

The final microstructure and the associated device stability achieved with NP oxidation
depends strongly on the relative humidity during NP deposition. Therefore, in
order to develop a reliable fabrication process, mitigating
day-to-day variation in RH of the ambient air and to build
understanding of the critical RH necessary for stabilization, a
new set of samples were prepared in a more controlled environment
using commercial dry synthetic air coupled with custom-built
humidifier (bubbler). The deposition with dry synthetic air
resulted in unstable samples with the corresponding SEM
micrographs [shown in \ref{bose1}(f)] indicating a slightly more
coalesced morphology in comparison to the RH $\sim$ 80\% ambient
air in \ref{bose1}(d). Although the microstructure has only subtle
differences, these differences become very important as the NP
system is poised near the percolation threshold and nanoscale
changes can promote(inhibit) inter-NP atomic-switch formation.

\subsection{Current surge protection}

\begin{figure*}[!t]
\centering
\includegraphics[width=15 cm ]{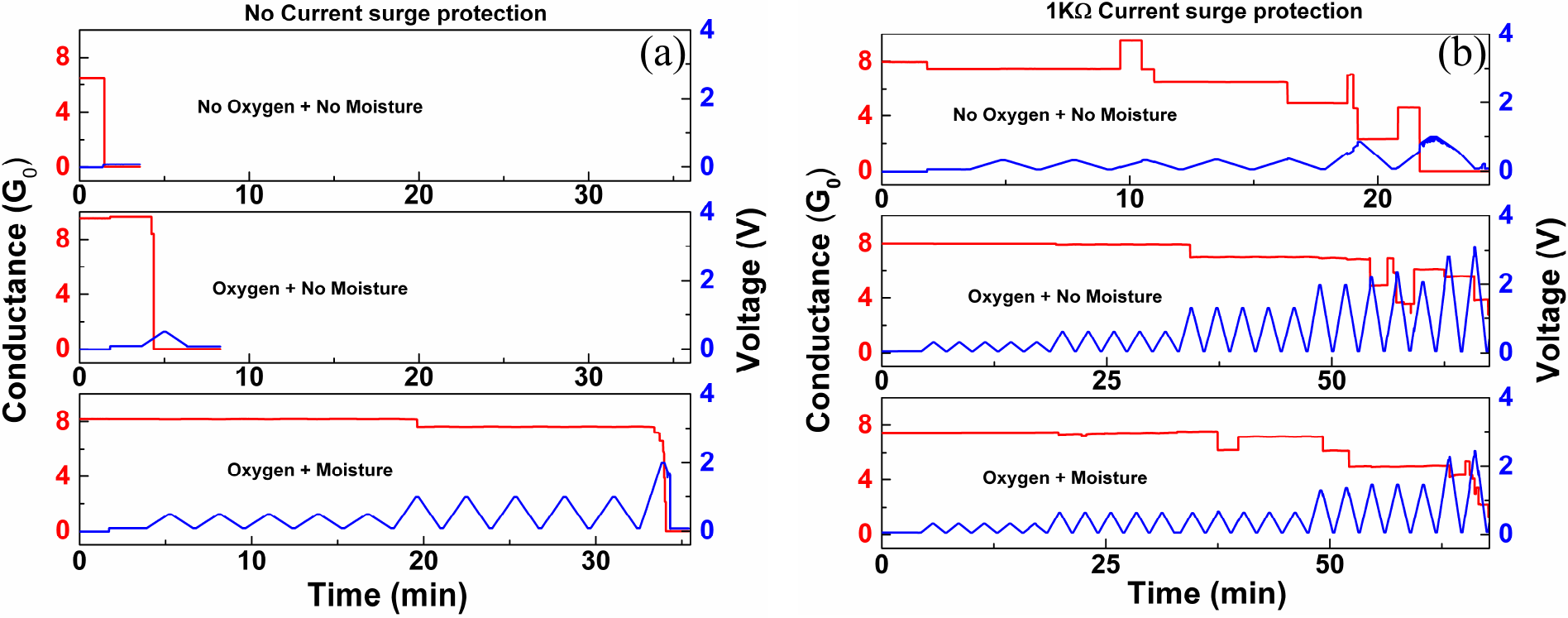}
\caption {Samples prepared with $P_{dep}= 10 \mu$Torr and
sub-optimum moisture content (RH $\sim$ 55\%) can be partially
stabilized with a current-surge protection in the form of a
1k$\Omega$ in-line resistor. The top panels in (a) and (b) show
the samples prepared without oxidation or moisture are unstable
even with the current surge protection. The samples become open
circuit ($\emph{G}$ $\rightarrow$ 0) on application of 0.1V (no
resistor) and 1V (with resistor). The middle panel in (a) shows
that samples prepared with dry oxygen (no moisture) are only
slightly more stable as $G$ $\rightarrow$ 0 at 1V (no resistor).
The equivalent sample with in-line resistor showed stability in
the data shown here but $\emph{G}$ $\rightarrow$ 0 in the next
measurement (not shown here). The bottom panels show that samples
prepared in oxygen and partial humidity ($\sim$ 55 $\%$) are more
stable than those in the middle panels (no moisture). The sample
prepared with oxygen + moisture and measured with 1k$\Omega$
resistor showing $\emph{G}$ switching for several weeks.}

\label{bose5}

\end{figure*}

The variation in conductance during the first voltage sweeps
applied to the new series of samples are shown in
Fig.~\ref{bose5}. The top panel of Fig.~\ref{bose5}(a) shows that
the atomic switch networks prepared at BP with no oxygen and no
moisture are disconnected on application of very small voltages of
0.1V. Introduction of dry air at $P_{dep}= 10 \mu$Torr, no
moisture [Fig.~\ref{bose5}(a) middle panel] leads to samples that
can only sustain small voltage sweeps. Samples prepared with the
same $P_{dep}= 10 \mu$Torr and higher RH $\sim$ 55\% are more
stable but still do not show sustained reconnections
($G_{\uparrow}$) and become disconnected at $\sim$2V, indicating
that these samples are only partially stabilized.  This provides
an opportunity to demonstrate an additional method for stabilizing
the switching behaviour in these devices.

Fig.~\ref{bose5}(b) shows that the devices show more stable
conductance switching when measured with a current-limiting
resistor (1k$\Omega$) in series with the device. The series
resistor limits the maximum allowed current flowing through the
percolating network and hence prevents the destruction of the key
connections via electromigration. The sample prepared with
$P_{dep}= 10 \mu$Torr but without moisture and measured with the
series resistor survived the voltage sweeps [Fig.~\ref{bose5}(b)
middle panel], but got disconnected in the next measurement (not
shown here). In contrast, the sample prepared with oxidation with
RH $\sim$ 55\% and measured with the in-line resistor was stable
for several weeks. Further samples prepared in synthetic air with
a higher RH ($\sim$60\%) exhibit stable (i.e. for months)
switching behaviour as in Fig.~\ref{bose2}, \emph{without} a
current-limiting resistor. This indicates that oxidation with a
critical amount of moisture RH ($\geq$60\%) creates a
microstructure which incorporates a robust current-limiting
resistor backbone, and thus do not require additional in-series
resistor protection.

\begin{figure}[!t]
\centering
\includegraphics[width=10 cm ]{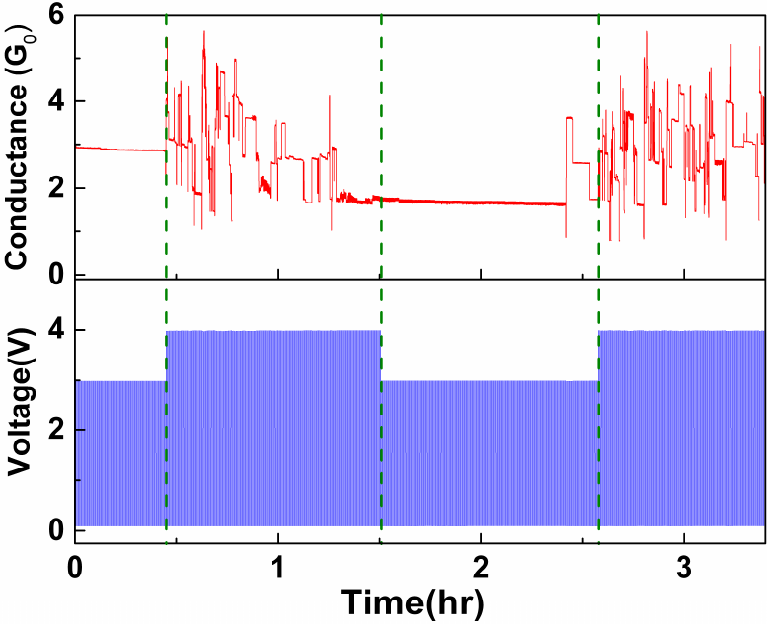}
\caption {Further conductance data of the sample described in
Fig.~\ref{bose2} showing that switching in these NP assemblies
requires application of a minimum voltage-stimulus. The 3V pulses
generate very few switching events, whereas  4V pulses trigger
multitude of switching events.}

\label{bose6}

\end{figure}

\subsection{Fabrication summary}
As discussed above, the crucial fabrication parameters for ASN stability are $P_{dep}$ and relative humidity RH\%. $P_{dep}$ was varied between BP (6 $\mu$Torr) and 600 $\mu$Torr, and RH\% was varied from completely dry (0\%) to nearly saturated (80\%). The optimal fabrication parameters for stability of these Sn cluster devices are $P_{dep}$ $=$ 10-50 $\mu$Torr and relative humidity RH $=$ 60-80\% when an in-series resistor of 1k$\Omega$ is used for current surge protection.

\section{Switching mechanism and dynamics}
\label{Dynamics}

To understand the physical process underlying the switching
mechanism, we present further voltage and time dependent studies.
Fig.~\ref{bose6} presents a segment of data acquired during the
long sequence of measurements on the sample used to obtain the
data in Fig.~\ref{bose2} (ambient air, $P_{dep} = 10\mu$Torr, RH
$\sim$ 80\%) showing that a critical voltage (or equivalently,
electric field) is required to activate the switching process. The
switching dynamics is voltage polarity independent, with negative
V pulses (not shown here) showing exactly the same switching
dynamics as positive V pulses. This polarity-independence allows
us to eliminate other possible switching mechanisms such as
Coulomb charging and electrochemical redox
reactions\cite{Avizienis2012}, and further substantiates the
electric-field and current induced switching mechanism described
here. As shown in Fig.~\ref{bose6}, the pulses with amplitude 3V
cause almost no switching events, whereas 4V pulses induce
multiple stochastic switching events. The inherently probabilistic
nature of the synaptic connections are clearly visible in the
snapshots shown in Fig.~\ref{bose2}, where stimulus near the
threshold voltage induces less than one switching event per pulse.
Such stochastic or probabilistic dynamics of the synapses are
integral to the functioning of the biological brain: e.g. the
opening and closing of synaptic ion channels and associated
transmission of neurotransmitter molecules is inherently
stochastic\cite{Rolls2010} and is understood to be critical for
noise-filtering\cite{Faisal2008}, signal transmission
\cite{Tuma2016} and reward-modulated Hebbian
learning\cite{Hoerzer2014}. The existence of a critical stimulus
strength (electric field here) for the stochastic formation
(annihilation) of atomic-scale wires in tunnel gaps in the network
is consistent with EFIE/EFISD (electromigration) mechanisms and
provides a unique global control over the synaptic network
reconfiguration.

\begin{figure*}[!t]
\centering
\includegraphics[width=15 cm ]{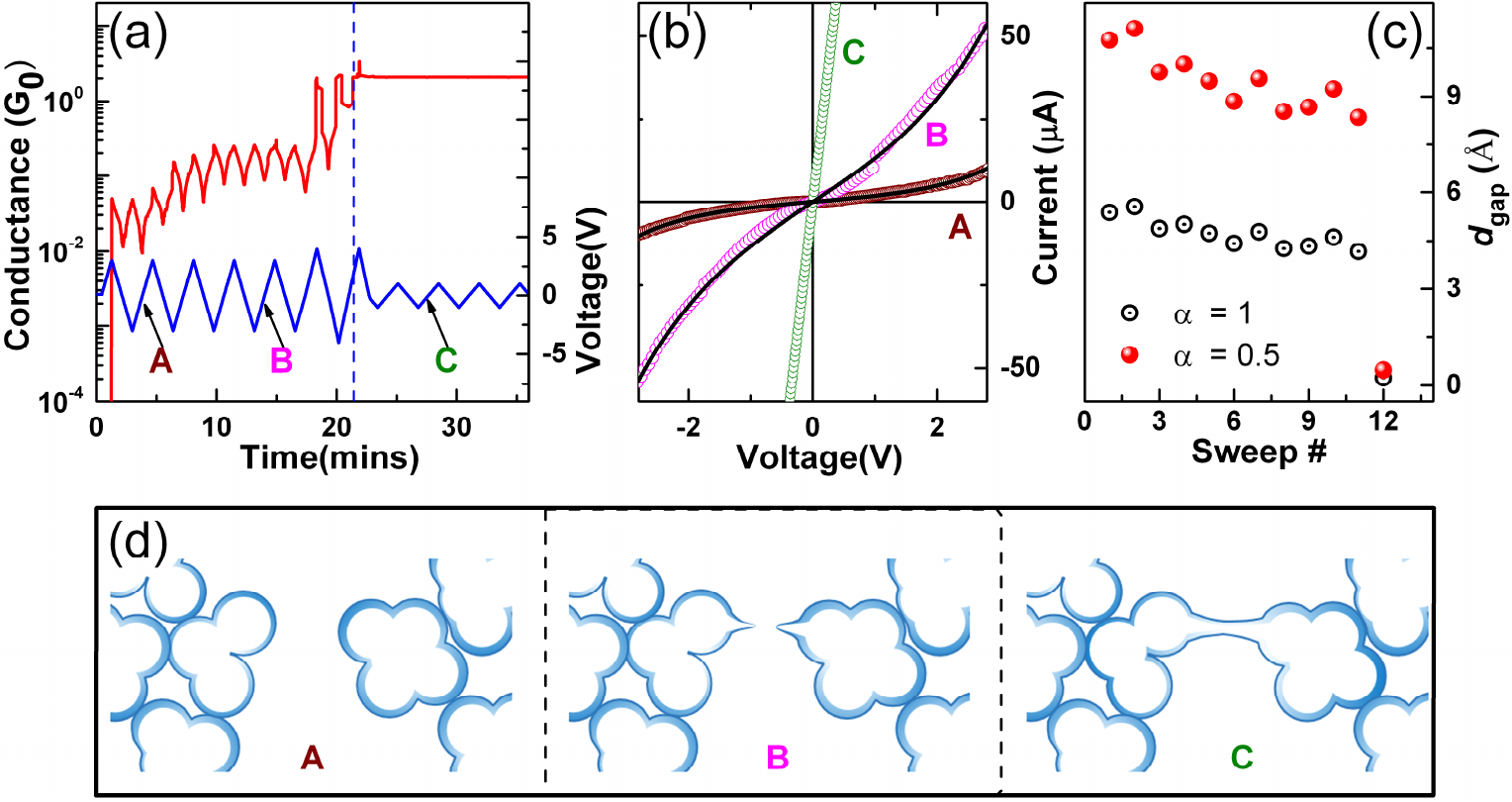}
\caption {(a--b) Slow bipolar voltage sweeps were applied to the
sample described in Fig.~\ref{bose2} when $\emph{G}$ $=$ 0. The
tunneling behaviour is evident from the non-linear current-voltage
\emph{I}(\emph{V}) characteristics e.g. at points A and B.
Successive voltage-sweeps increase $\emph{G}$ by two orders of
magnitude before the jump-to-contact and $G$ $\rightarrow$ 2G$_0$
at $\sim$22 min. This newly formed connection is then stable under
further voltage sweeps and shows the expected linear
\emph{I}(\emph{V}) characteristics. The solid lines in panel (b)
are fits to Simmons' tunneling model (\ref{eq:Simm01}). (c) The
calculated tunnel gap $\emph{d}_{gap}$ decreases monotonically
before an ohmic connection is formed at sweep 12. The
\emph{I}(\emph{V}) is ohmic after that (e.g. position C). (d)
Schematic depicting the gradual formation of atomic wire in one of
the tunnel barrier.}

\label{bose7}

\end{figure*}

To further validate the model and estimate the effective tunnel
barrier parameters associated with the gap in which the atomic
wires are formed, pulsed voltage measurements were stopped when
the device became open circuit (i.e. $G < 10^{-5}$$G_0$). A series
of slow bipolar voltage sweeps were then applied which showed
non-linear current-voltage \emph{I}(\emph{V}) characteristics as
in Fig.~\ref{bose7}(a--b). On the 1$^{st}$ voltage sweep, \emph{G}
jumps from $< 10^{-5}$$G_0$ to $\sim 10^{-2}G_0$, which
corresponds to the formation of a tunnel gap which is sufficiently
small to allow a measurable current to flow. The corresponding
tunneling current through a non-ideal potential barrier with
height $\Phi_{B}$ and width $\emph{d}$ is\cite{Simmons1963}:

\begin{equation}\label{eq:Simm01}
\begin{split}
I\propto{}(\Phi_B-eV/2)\exp
\Big[-\frac{2(2m)^{1/2}}{\hbar}\alpha(\Phi_B-eV/2)^{1/2}d \Big]\\
-(\Phi_B+eV/2)\exp
\Big[-\frac{2(2m)^{1/2}}{\hbar}\alpha(\Phi_B+eV/2)^{1/2}d \Big]
\end{split}
\end{equation}

with $m$ being the free electron mass and $\alpha$ being an
adjustable parameter representing the non-ideal character of the
tunneling barrier and effective electron mass. Fig.~\ref{bose7}(b)
shows representative \emph{I}(\emph{V}) characteristics for the
voltage sweeps marked A, B and C in Fig.~\ref{bose7}(a) along with
the fits to (\ref{eq:Simm01}) (solid lines). The associated
barrier width \emph{d}, with calculated barrier height $\Phi_{B}$
$\sim$ 2 eV, decreases monotonically with the sweep number \#,
when either an ideal ($\alpha =$ 1) or highly non-ideal ($\alpha
=$ 0.5) barrier is assumed. Interestingly the resultant electric
field exceeds the $\sim$ 1 Vnm$^{-1}$ threshold for EFISD but
remains lower than the $\sim$ 25 Vnm$^{-1}$ required for
EFIE\cite{Sattar2013,Olsen2012}. As shown schematically in
Fig.~\ref{bose7}(d), the narrowing of the tunnel gap under the
influence of the electric field continues, until after about 20
minutes an atomic scale wire closes the tunnel gap, i.e. a ``jump
to contact" occurs\cite{Agrait2003} leading to a conductance $G =
2 G_0$. These atomic-scale wires are similar to those formed in
mechanically controlled break junctions (MCBJs)\cite{Agrait2003}.
The wire breaks and re-forms a couple of times
[Fig.~\ref{bose7}(a)] and then is observed to be completely stable
when subjected to further voltage sweeps. The Ohmic conductance is
marked by linear \emph{I}(\emph{V}) behavior depicted in curve C
of Fig.~\ref{bose7}(b). Such conductance modulation with
successive stimulus (electric-field) is one of the key
requirements for synaptic learning capability in neuromorphic
systems\cite{Fostner2015,Hasegawa2010}, and is similar to the
sensory memory reported in Ref.\cite{ohno2011a}.

\begin{figure}[!t]
\centering
\includegraphics[width=10 cm ]{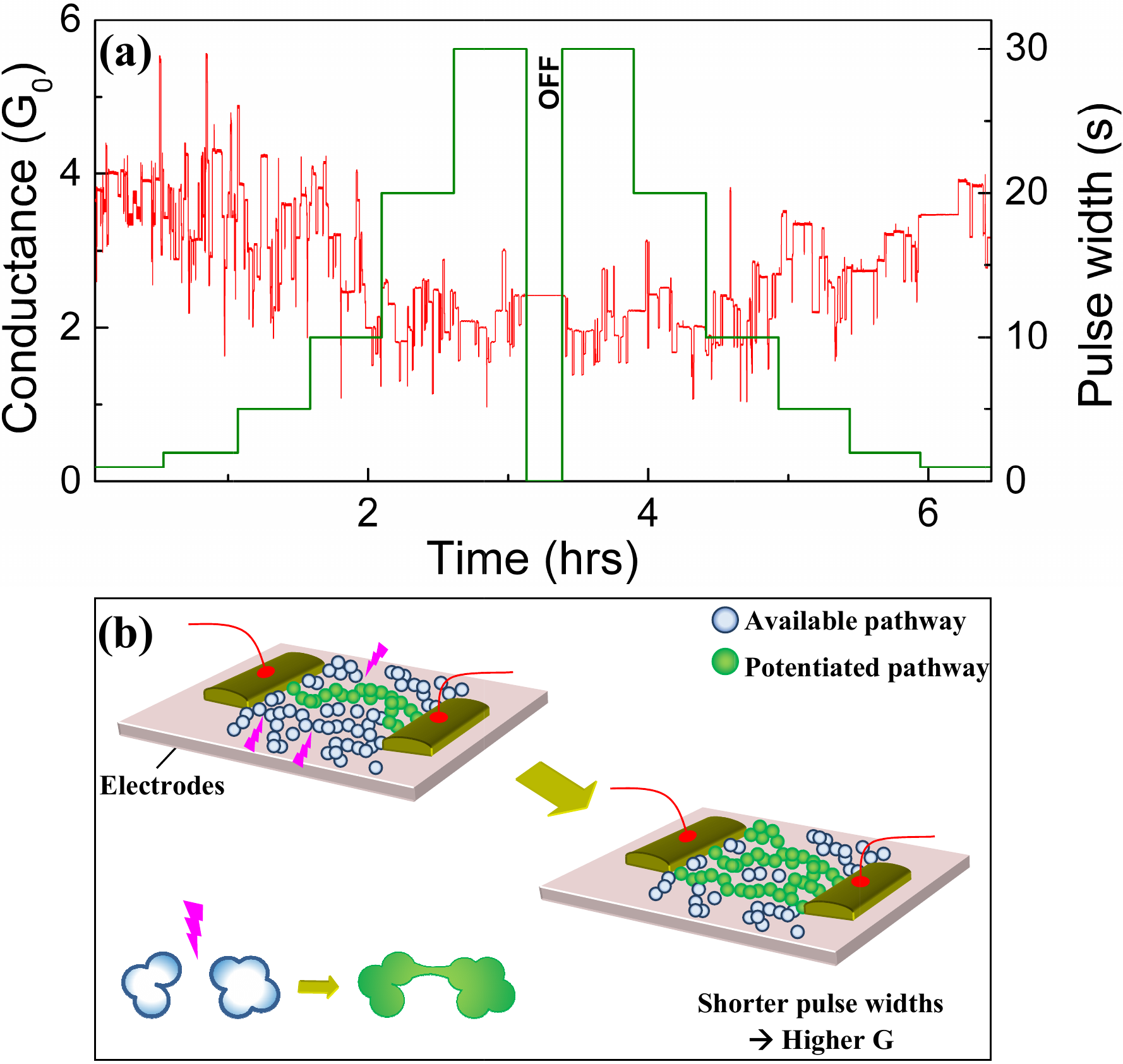}
\caption {(a) Synaptic plasticity dependence on stimulus frequency
is depicted in switching behaviour in response to a sequence of
voltage pulses with fixed $V_p$=4V and variable pulse widths
$\tau_p$ (1 -- 30s). The conductance remains unaltered for read
voltages of 0.1 V as seen in the flat section (OFF) in middle of
the sequence. (b) Schematic depiction of representative synaptic
pathways with only very few pathways shown for clarity. The real
device ASN is much more complex. Shorter pulse widths leads to
electric-field induced connections as described in
Fig.~\ref{bose7}. Formation of additional atomic-wire connections
(one depicted here) cause more potentiated synaptic pathways and
thus higher conductance as seen in panel (a).}

\label{bose8}

\end{figure}

Realization of neuromorphic behavior in these ASNs also requires a
scheme to modify the density of potentiated synaptic
pathways\cite{Fostner2015}. In Fig.~\ref{bose8} we show such
stimulus frequency dependent potentiation. Square voltage pulses
with fixed $V_p$=4V (just above threshold voltage) with various pulse widths
$\tau_p$ (1, 2, 5, 10, 20, and 30 s) are applied successively for
30 mins each (the green lines depict the pulse-width). Longer
$\tau_p$ (slow pulses) leads to lower conductance whereas shorter
$\tau_p$ (faster pulses) leads to additional formation of synaptic
pathways as schematically depicted in Fig.~\ref{bose8}(b),
resulting in higher $G$. This variation can be understood in the
light of the electric field-induced reconnections dominating over
the electromigration induced disconnection of atomic wires for
shorter $\tau_p$. Such stimulus rate dependent reconfiguration of
network connectivity can be modeled as short-term to long-term
memory conversion\cite{ohno2011,Chang2011}.

\section{Conclusion}

In summary, we have demonstrated a unique approach for realization
of self-assembled atomic switch networks with stimulus induced
control of the synaptic configuration reflected in the device
conductance. By controlling oxidation and humidity during NP
deposition, nanoparticle coalescence is inhibited resulting in
stochastic switching that is stable over several months. The
atomic-wire formation in these \textit{oxidised} nanostructures is
very surprising and detailed modeling~\cite{Onofrio2015} of the
atomistic mechanisms~\cite{Xiang2009,Olsen2012,Sattar2013} in the
presence of oxides~\cite{Sutter2014,Lee1998} is required, as is atomic scale modelling of
the effect of humidity~\cite{Cho2005} on the oxidation process.

We have also highlighted the stochastic nature of the switching
mechanism together with the stability of the distribution of switching events - these reflect
inherently complex network dynamics that are a key requirement for
neuromorphic applications. The next stage of this research will be to build devices with multiple contacts and to demonstrate that the networks exhibit the required network dynamics. The precise requirements are different for different applications\cite{Burr2017, Nawrocki2016}, but for RC include distributed spatio-temporal dynamics, recurrency, higher harmonic generation, switching speed, network size and type of connectivity\cite{Mantas2009, Kulkarni2012a, Konkoli2014, Avizienis2012,Busing2010, Demis2016}.

More generally, these complex percolating
structures mimic some of the features of biological neural networks and synaptic structures and
so could provide a foundation for a range of future neuromorphic architectures. Future work will
focus on utilizing these structures to implement previously suggested algorithms~\cite{Konkoli2014,Kulkarni2012a}
and hence to demonstrate utility in key applications.

\section*{Acknowledgment}
The technical support of G. MacDonald and G. Graham are
acknowledged.

\bibliographystyle{IEEEtran}

\end{document}